# *On the Alleged Use of Keplerian Telescopes in Naples in the 1610s*

By Paolo Del Santo

Museo Galileo: Institute and Museum of History of Science – Florence (Italy)

**Abstract**

The alleged use of Keplerian telescopes by Fabio Colonna (c. 1567 – 1640), in Naples, since as early as October 1614, as claimed in some recent papers, is shown to be in fact untenable and due to a misconception.

At the 37[th] Annual Conference of the Società Italiana degli Storici della Fisica e dell'Astronomia (Italian Society of the Historians of Physics and Astronomy, SISFA), held in Bari in September 2017, Mauro Gargano gave a talk entitled *Della Porta, Colonna e Fontana e le prime osservazioni astronomiche a Napoli*. This contribution, which appeared in the *Proceedings* of the Conference (Gargano, 2019a), was then followed by a paper —actually, very close to the former, which, in turn, is very close to a previous one (Gargano, 2017)— published, in the same year, in the *Journal of Astronomical History and Heritage* (Gargano, 2019b). In both of these writings, Gargano claimed that the Neapolitan naturalist Fabio Colonna (c. 1567 – 1640), member of the Accademia dei Lincei, would have made observations with a so-called "Keplerian" or "astronomical" telescope (i.e. with converging eyepiece), in the early autumn of 1614.

The events related to the birth and development of Keplerian telescope —named after Johannes Kepler, who theorised this optical configuration in his *Dioptrice*, published in 1611— are complex and not well-documented, and their examination lies beyond the scope of this paper. However, from a historiographical point of view, the news that someone was already using such a configuration since as early as autumn 1614 for astronomical observations, if true, would have not negligible implications, which Gargano himself does not seem to realize. As a matter of fact, it would mean to backdate of a couple of years the *terminus ante quem* for the actual realisation of the very first Keplerian telescope (the most ancient known written source concerning a concrete example of this optical combination is a document of 1616, kept at the Tiroler Landesmuseum Ferdinandeum, in Innsbruck; Daxecker, 2004: 14), and of at least fifteen years its introduction in the observational praxis.

Gargano's claim is essentially based upon a passage of a letter that Fabio Colonna wrote to Galileo Galilei on the occasion of the solar eclipse of 3 October 1614. The letter contains six sketches depicting the Sun's surface, partially eclipsed by the Moon, with the sunspots. The observation was made by projecting the solar image through the telescope onto a sheet of paper, a method widely used, among others, by Galileo himself, and first suggested by his disciple Benedetto Castelli (Galileo, 1613: 52). In the above-mentioned letter, Colonna apologises for the poor quality of his drawings, and invites Galileo to salvage as much as possible and to "turn them right side up, since they came out from the telescope inverted [alla riversa]" (Colonna, 1614).

That is how Gargano (2019b: 54) has interpreted Colonna's words:

Upon reading this letter it is evident that Colonna used a telescope and not a Galilean spyglass.[1] Therefore, this was the first astronomical observation made from Naples using a Keplerian-like refractor.

But did Colonna really use a Keplerian telescope in his astronomical observations? As we shall see, the answer to this question is definitely no! Indeed, as it is well known and easy to prove, in the direct observation, a Galilean telescope provides upright images and a Keplerian telescope upside-down ones, but, if used in projection, a Galilean telescope gives upside-down images and a Keplerian telescope upright ones. Gargano therefore, seems not only to be unfamiliar with geometrical optics, but with the early history of the telescope as well. Otherwise, he would know the following passage from the renowned work by Galileo *Istoria e dimostrazioni intorno alle macchie solari e loro accidenti*, published in March 1613, i.e. one year and a half before Colonna's letter:

> It should be noted next that [using the telescope by projection] the spots exit the tube inverted and located opposite to where they are on the Sun: that is, the spots on the right come out on the left side, and the higher ones lower, because the rays intersect each other inside the tube before they emerge from the concave glass. But because we draw them on a surface facing the Sun, when turning back toward the Sun, we hold the drawing up to our eyes, the side on which we drew no longer faces the Sun but is instead turned away from it, and therefore the parts of the drawing on the right-hand side are already in their proper place again, corresponding to the right side of the Sun, and the left ones on the left, such that one only has to invert the upper and lower ones. Therefore, turning the paper over and thus making the top the bottom, and looking through the paper while facing the light, one observes the spots as they should be, as if we were looking directly at the Sun. And in this appearance they must be traced and inscribed on another sheet in order to have them correctly positioned. (Galilei, 1613: 53; translation from Reeves and Van Helden, 2010: 127).

So, if we did not know the laws of geometrical optics, in the light of this detailed description, following Gargano's belief, we should be forced to think that Galileo was wrong, or that Galileo himself had already abandoned the optical combination named after him, in favour of the Keplerian one, at the beginning of the 1610s!

Besides, Gargano seems also not to know the passage of the *Rosa Ursina* in which the German Jesuit Christoph Scheiner describes the projection technique by using both a concave (diverging) lens, and, as an alternative to it, a convex (converging) one (Scheiner, 1630: 129$^v$, 130$^r$). Therein, Scheiner correctly states that the telescope projects upside-down and right-side up images, respectively.

---

1 It is worth noting, incidentally, that Gargano seems here to use improperly the terms "telescope" and "spyglass". This erroneous use of the two terms is even clearer in the same passage of the original Italian version of the paper, in which Gargano (2019a: 271) uses "telescopio" and "cannocchiale", to indicate a Keplerian telescope and a Galilean (or Dutch) telescope, i.e. with diverging eyepiece, respectively. However, the two Italian terms were originally synonyms, while nowadays "cannocchiale" survives, with a pretty ambiguous meaning, almost exclusively in the everyday language, and it is usually used to indicate a refracting telescope, especially if modest in size (so, only in the latter acceptation, namely in the meaning of handle telescope, it can be translated into English as "spyglass"). In any case, in no way the possible semantic difference between the two terms lies in the fact that a "telescopio", without any attribute (like "telescope" in the English version as well) would produce upside-down images, and a "cannocchiale" erected ones. On the other hand, the inadequacy of Gargano's linguistic baggage shines trough other passages too. For instance, talking about the lack of biographical information concerning Francesco Fontana, Gargano quotes Crasso (1666: 298), who, referring to Fontana's life, says "oscuro Fato d'un'Huomo illustre", a sentence which Gargano (2019b: 57, endnote n. 6) translates as "unclear Fate of a famous man". However, Crasso is here using the word "oscuro" not with the meaning of "scarcely known" or "not well-documented", but with the nowadays rather obsolete meaning of "gloomy" or "sombre".

One could argue that the expression "alla riversa", used by Colonna in his letter to Galileo, might refer not to the vertical (up-down) inversion, but to the horizontal (left-right) one. Actually, "riversa" is an Italian archaic term for the modern "rovescio" or "rovescia", which, used in the locutions "al rovescio" or "alla rovescia", have the generic meaning of "upside down", "wrong side up", "wrong way round" or even "back to front" (inside out). However, even though the optical paths of the Galilean and the Keplerian combination are completely different, when used by projection, both produce mirrored images, namely, if one looks the screen from the side of the eyepiece, left and right are inverted. This is why, in order to get all orientations correct, both Galileo, as we have seen above, and Scheiner (1630: 129$^v$, 130$^r$) recommend transposing the projected image onto the opposite surface of the sheet, tracing it against the light. Hence, the two optical configurations are indistinguishable on this account, and therefore, even if Colonna really meant a mirror-like inversion of the image, this circumstance does not prove anything about the type of optical combination, Galilean or Keplerian, he used.

We have already touched upon the implications of Gargano's reconstruction on the chronology of the development of the Keplerian telescope, but he goes further: he claims that the new optical configuration "did not derive from Kepler's studies or those of Fontana, but was the result of the combined theoretical and practical skills of Della Porta and Colonna" (Gargano, 2019: 54), who therefore, in Gargano's opinion, would be the true originators of the Keplerian telescope. Gargano's statement is based on a letter from Della Porta to Galileo, dated 26 September 1614, i.e. exactly one week before the aforementioned letter from Colonna to Galileo of the 3$^{rd}$ of October. In his letter, Della Porta informs Galileo that he is working, together with Colonna, to realize "a new kind of telescope, that will increase a hundredfold the performance of the usual ones" (Della Porta, 1614). However, Gargano (2019: 54) takes for granted that Della Porta and Colonna were working on the development of a Keplerian telescope —which, as we shall see later, is more than unlikely, and, in any case, unproven— and that this new, unspecified instrument was the one used by Colonna himself to observe the eclipse just a week later. As a matter of fact, neither in the letter to Galileo of the 3$^{rd}$ of October nor elsewhere in the following months, Colonna claims he is using a new kind of telescope, different from those (Galilean) that he usually used in his observations. Nor is there evidence that suggests such a hypothesis. Frankly, I find rather odd that Colonna was satisfied with testing (furthermore, on his own without Della Porta, his partner in the project) such an innovative instrument just with one observation, made returning home hurriedly, between two judicial hearing in the courthouse of Naples, where he practised law for a living (Colonna, 1614). On the other hand, as far as we know, Della Porta himself, who was to die a few months later, does not mention any more his mysterious telescope, and, in all likelihood, his project led to no result.

In my opinion, the irrefutable evidence that Colonna did not use a Keplerian telescope in 1614 is a letter —which Gargano evidently does not know (indeed, neglected by most historians of the telescope)— from Colonna himself to Federeico Cesi, one of the founders of the Accademia dei Lincei. The letter, dated 19 September 1626, contains a reference to a small

telescope, made by the Neapolitan optician Francesco Fontana, which undoubtedly had a converging eyepiece:

> This friend [Francesco Fontana] has also invented a […] telescope, long just a palm,[2] which shows objects upside-down, but magnifies them very much, and, what is most remarkable, it shows objects so near that those which are as far away as a musket shot are seen close to the eyes; I could not understand yet, because I cannot attend to that, due to my job, how it can do that, since two convex lenses show objects more distant and smaller than they actually are, but more sharply; when I get around to going to him, I will examine it, and I will obtain one to His Excellency. (Colonna, 1626)

That passage shows conclusively that —twelve years after his letter to Galileo— Colonna still could not understand, both theoretically and practically, the way two convex lenses can form a telescope!

In conclusion, there is no doubt that the blunder committed by Gargano is due to his lack of knowledge of the primary and secondary sources about the early history of the telescope (and of the laws of the geometrical optics as well), a history which, however, he aspires to rewrite. Nevertheless, Gargano probably is perpetrator and victim, at the same time, of a narration, often based more on the imagination than on historiographic sources (although, to be fair to him, not with equal sensationalism), chiefly due to Paolo Molaro and Pierluigi Selvelli, who, in several papers published over the last decade, have conceived and promoted: 1) The (wrong) idea that the telescope depicted in the *The Sense of Sigh,* painted by Jan Brueghel the Elder and Peter Paul Rubens in 1617, is a Keplerian one, and therefore, that this optical combination was already pretty common at that time (Molaro and Selvelli, 20: 331-32); 2) The (wrong) idea —based on the (wrong) assumption that "[t]here are no apparent reasons to question Father Zupus's declaration to have used Fontana's [Keplerian] telescope in 1614"— that "the year of 1608 does not seem so implausible as the birthdate of Fontana's [Keplerian] telescope" (Molaro, 2017b: 284-86; see also Molaro, 2017a: 227); 3) The (wrong) idea, consistent with the previous ones, that, as early as around 1615, Fontana already enjoyed such a reputation as optician that one of the greatest painter of his time, José de Ribera, would have portrayed him in his painting *Allegory of Sight* (Molaro, 2017: 284-86). Well, if one believes all those things, probably, one will consider plausible, even likely, that, in Naples, in autumn 1614, someone was already making astronomical observations by means of a Keplerian telescope...

---

2  Until 1840, the Neapolitan palm was 0.263670 m.